\documentclass[aps,prl,reprint,superscriptaddress]{revtex4-1}
\usepackage{graphicx}
\usepackage{epstopdf,gensymb}
\usepackage{color}
\usepackage{subfigure}
\usepackage{dcolumn}
\usepackage{bm}
\usepackage{float}

\begin{document}
\title{Observation of Self-Sustaining Relativistic Ionization Wave Launched by Sheath Field}

\author{M. W. McCormick}
\affiliation{Center for High Energy Density Science, The University of Texas, Austin, Texas 78712, USA}

\author{A. V. Arefiev}
\affiliation{Institute for Fusion Studies, The University of Texas, Austin, Texas 78712, USA}

\author{H. J. Quevedo}
\affiliation{Center for High Energy Density Science, The University of Texas, Austin, Texas 78712, USA}

\author{Roger D. Bengtson}
\affiliation{Center for High Energy Density Science, The University of Texas, Austin, Texas 78712, USA}

\author{T. Ditmire}
\affiliation{Center for High Energy Density Science, The University of Texas, Austin, Texas 78712, USA}

\date{\today}
\begin{abstract}
We present experimental evidence supported by simulations of a relativistic ionization wave launched into surrounding gas by the sheath field of a plasma filament with high energy electrons. Such filament is created by irradiating a clustering gas jet with a short pulse laser ($\sim$115 fs) at a peak intensity of $5 \times 10^{17}$ W/cm$^2$. We observe an ionization wave propagating radially through the gas for about 2 ps at 0.2-0.5 $c$ after the laser has passed, doubling the initial radius of the filament. The gas is ionized by the sheath field, while the longevity of the wave is explained by a moving field structure that traps the high energy electrons near the boundary, maintaining a strong sheath field despite the significant expansion of the plasma.
\end{abstract}

\pacs{}
\maketitle


A sheath electric field is a ubiquitous phenomenon caused by electron motion that occurs at the plasma edge. It is employed in applications ranging from plasma discharges~\cite{Song1991,Stenzel2011} to ion acceleration~\cite{Snavely2000,Wilks2001}. In this paper, we present experimental evidence supported by simulations of a collisionless self-sustaining relativistic ionization wave that is launched into a surrounding un-ionized gas by a plasma sheath field. 

Suitable experimental conditions are achieved by irradiating a supersonic clustering argon gas jet with a moderate intensity laser pulse ($5 \times 10^{17}$ W/cm$^2$). Enhanced absorption of the laser energy by such jet makes it possible to create a plasma with high energy electrons~\cite{Ditmire1996}. The expansion of such plasmas have been used to study phenomena ranging from nonlocal heat transport~\cite{Ditmire1998} to formation of electrostatic shocks~\cite{Nilson2009} and blast waves~\cite{Ditmire2000}. The corresponding time-scale is in the range of tens of ps, because these phenomena either involve ion motion or electron collisions. In what follows, we focus on much faster phenomenon. We present direct measurements of a relativistic velocity ionization wave, launched by the radial sheath field of a laser-generated plasma with high energy electrons, that is sustained for up to 2 ps. The measured radial velocity of the wave after the laser pulse is 0.2-0.5 $c$, causing the plasma radius to double on a ps time scale. Our relatively short laser pulse (115 fs) makes it possible to clearly distinguish energy deposition into the plasma from the propagation of the ionization wave that follows.

The measured increase of the plasma radius is clearly too fast to be attributed to hydrodynamic motion, and it is even too fast to be explained by free-streaming electrons ionizing the surrounding gas via impact ionization. Indeed, the velocity of these electrons has to be comparable to the speed of the wave front, $v_e \approx 0.5$ $c$, which leads to a collisional time $\tau_e \gg 10$~ps at $3 \times 10^{18}$~cm$^{-3}$ gas densities~\cite{Talukder2008}. This is much too long for the observed ionization. In this paper, we also present particle-in-cell (PIC) simulations that reveal a new collisionless mechanism that can launch and maintain the observed fast ionization front. We show that a hot electron minority produced by the laser-cluster interaction generates a strong sheath electric field at the edge of the plasma that can then ionize the surrounding gas. The electrons produced from this ionization alter the sheath field to create a narrow potential well that traps the energetic electrons. These electrons then remain bunched near the boundary, maintaining the strong sheath field during the plasma expansion.

\begin{figure}[tb]
\centering
\includegraphics[height=.22\textheight]{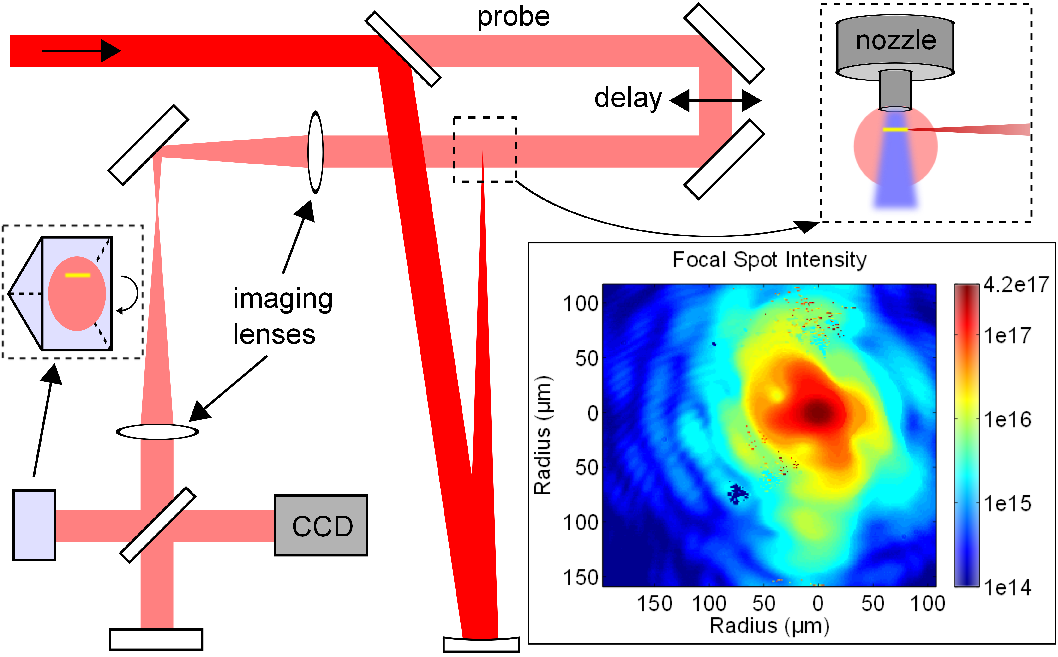}
\caption{(Color online) Schematic layout of the experimental setup. The main beam is focused under the nozzle, with the spatial profile shown in the inset, to generate the plasma. The probe beam is perpendicular to the main beam and is relay imaged into a Michelson interferometer with a right-angle prism in one arm for electron density measurements.}
\label{Figure1}
\end{figure}

In our experiments, we used the GHOST laser~\cite{Hays2007}. The front end uses optical parametric chirped pulse amplification (OPCPA) to produce high gain with excellent contrast, measured to be at least $10^7$, while the final stage uses mixed glass amplification to keep the spectrum broad and allow for compression to 115 fs FWHM at 1057 nm. In these experiments, most shots were around 1 J on target. A schematic view of the experimental setup is shown in Fig.~\ref{Figure1}. The laser is focused using an $f/23$ spherical mirror, producing a spot size of 35 $\mu$m FWHM under the output of the gas jet and a peak intensity of $5 \times 10^{17}$ W/cm$^2$. Supersonic Ar and He gas jets are produced by a $2.5\degree$ half angle nozzle backed by gas at 10.5 and 7 bar. Under these conditions, the Ar jet contains clusters with 5-10 nm average radius~\cite{Dorchies2003}, whereas the He jet consists only of monomers. The leakage of the pump beam through one mirror provided the short pulse probe beam, and an optical delay stage allowed for sub-ps time steps. The top part of the beam with the interaction information was flipped by the prism and interfered with the bottom reference part of the beam to measure the electron density interferometrically. The $f/\#$ of the lens after the interaction was selected so that it collected all the refracted probe light, as simulated by ray tracing.

\begin{figure}[tb]
\centering
\includegraphics[height=.29\textheight]{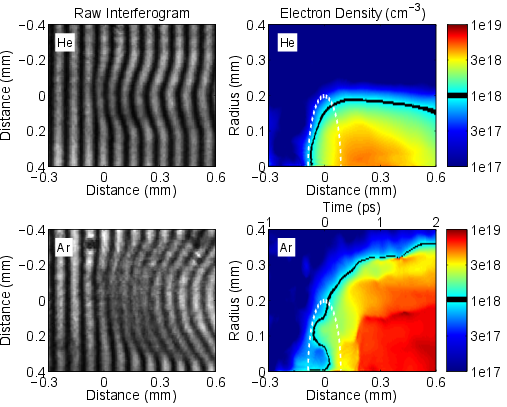}
\caption{(Color online) Interferometric images, cropped from originals to show more detail, for 10.5 bar helium and argon jets heated by the 115 fs laser pulse, with retrieved electron density profile after Abel inversion. The spatial axis is converted to a time axis as described in the text, with the white line corresponding to $I=10^{14}$ W/cm$^2$ of the main beam.}
\label{Figure2}
\end{figure}

The phase information from the interferometric images in Fig.~\ref{Figure2} is converted to an electron density profile along the laser axis using an Abel inversion technique. Since we are interested primarily in the electron densities near the plasma edge, irregularities near the axis of the plasma density plots, resulting from the probe traversing the plasma while it is quickly evolving, were ignored as they do not affect our measurements at large radii. The qualitative difference in ionization between the He and Ar jets is evident from the sample images shown in Fig.~\ref{Figure2}, where the ionization in He occurs only during the laser pulse, while the ionization in Ar continues long after the laser pulse has passed. The rising edge of the pulse starts ionizing the jet at least 200 fs before the peak laser intensity is reached, because the $10^{14}$ W/cm$^2$ intensity is sufficient to ionize the gas (a sech$^2$ temporal profile would start ionizing 300 fs before peak). This is consistent with the recovered electron density time history in He, where the radius of the plasma filament increases for 300 fs and then remains constant. The radius of the plasma filament (200 $\mu$m) significantly exceeds the FWHM of the focal spot (35 $\mu$m) because of the spatial lower intensity wings of the focal spot, shown in the inset of Fig.~\ref{Figure1}. Because of the lower barrier suppression ionization intensity for Ar (about 5 times lower for $Z=1$), the laser initially ionizes the gas to a slightly larger radius than in the He case. However, the ionization continues as the plasma expands to nearly 400 $\mu$m on a ps time scale after the laser has passed. This increase of the plasma radius in Ar is clearly caused by ionization, rather than expansion, since the interior electron density does not decrease.

The measured average atomic density in the gas jet for 10.5 bar backing pressure is about $3 \times 10^{18}$ cm$^{-3}$. This implies that the average ionization $Z_{av}$ is 1-2 during the first ps despite the fact that the laser field can ionize atoms to at least $Z \approx 6$ at our peak intensities. A likely reason for such low $Z_{av}$ in the early Ar plasma is that the density measured on a ps time scale is primarily the electron density produced by monomers. The electron density inside ionized clusters on a sub-ps time scale remains over-critical and contributes very little to the refractive index~\cite{Gao2012,Gao2013}.  At later times, when the plasma evolution is slower and the clusters have disassembled, the measured $Z_{av}$ inside the filament is about 6.

\begin{figure}[tb]
\centering
\includegraphics[height=.19\textheight]{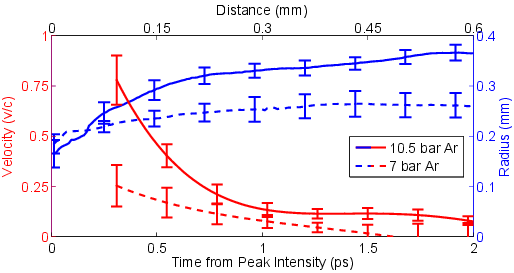}
\caption{(Color online) Radial position and instantaneous velocity of $10^{18}$ cm$^{-3}$ electron density contours for argon at 7 bar and 10.5 bar backing pressures. The velocity curves start after the end of the main pulse.}
\label{Figure3}
\end{figure}

While the interferometric images in Fig.~\ref{Figure2} show a 2D spatial snapshot of the plasma, a time history of the expansion can also be inferred. The undeflected fringes to the left show un-ionized gas (or clusters), and the right side shows plasma after the laser has passed. We note the following: (1) the Rayleigh length of the main beam is 2.5 mm, so the spatial profile variations along the laser axis are relatively small; (2) the plasma radius in later snapshots (not shown) is uniform; (3) the main beam is observed to move at approximately $c$ across the observation window when comparing ps-separated snapshots. Therefore, tracking a constant density contour near the radial edge of one snapshot measures the instantaneous expansion rate of the plasma. We fit a curve to a contour of $10^{18}$ cm$^{-3}$, which is chosen because it is near the radial edge and corresponds to near $Z = 1$ ionization, and the resulting slope yields the instantaneous velocity of the plasma boundary. Plots of the extracted contours and calculated velocity are shown in Fig.~\ref{Figure3}. The plasma expands at speeds which are a significant fraction of $c$ for over 1 ps after the influence of the laser is over. Figure~\ref{Figure3} also shows the $10^{18}$ cm$^{-3}$ density contours and corresponding instantaneous velocities for 7 bar backing pressure Ar. While not shown, we have also computed the average velocity by taking the radial difference of contours from separate laser snapshots and dividing by the time between the snapshots (1 ps minimum). The velocity measured by this method is comparable, but has larger error bars resulting from shot to shot variations in laser energy.

A comparison of the results at 10.5 and 7 bar shows that the instantaneous velocity and the maximum plasma radius decrease with the backing pressure. The cluster radius also decreases with the backing pressure~\cite{Hagena1987}, which indicates that the observed ionization wave is directly linked to the presence of large clusters. The fact that no ionization wave is observed in He adds further support to this conclusion. It is well known that laser ionization of clusters can generate energetic electrons with experimentally measured energies exceeding the ponderomotive potential~\cite{Shao1996,Fukuda2007,Zhang2012}. Large clusters in the Ar jet can thus generate a hot electron minority with energies well above 100 keV for our laser intensities either via inverse bremsstrahlung or collisionless heating mechanisms~\cite{Antonsen2005,Breizman2005,Arefiev2010}. Therefore, the key difference between He and Ar is that the laser generates a plasma with a hot electron population in Ar, whereas the He plasma contains none. The hot electrons are not able to leave the filament, as they are confined by an ambipolar sheath electric field generated at the edge of the plasma. If the field is sufficiently strong, it can directly ionize the surrounding gas, leading to plasma expansion. Since the mechanism is collisionless, its rate increases with electron energy, which makes it a good candidate to explain the observed fast ionization wave in the presence of energetic electrons.

\begin{figure}[tb]
\centering
\subfigure{\includegraphics[height=.15\textheight]{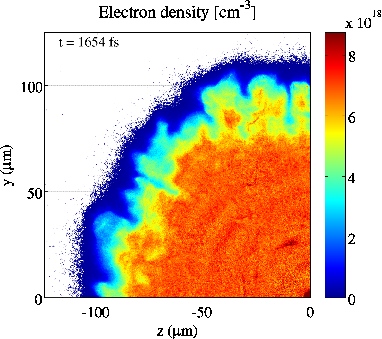}} 
\subfigure{\includegraphics[height=.15\textheight]{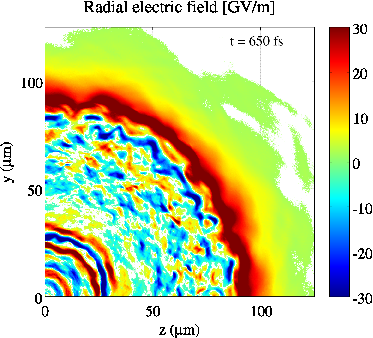}} 
\\
\subfigure{\includegraphics[height=.25\textheight]{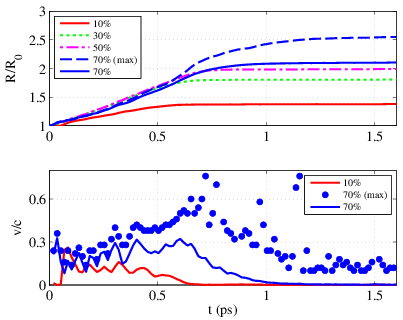}}
\caption{(Color online) Simulated electron density and radial electric field in Ar for an initial plasma radius $R_0 = 50$ $\mu$m. The lower two plots show the angle-averaged radius $R$ and velocity $v$ for different initial fractions of hot electrons. The maximum values of $R$ and $v$ for the 70\% case are also shown.}
\label{Figure55}
\end{figure}

In order to determine the feasibility of this hypothesis, we have performed 2D PIC simulations where a plasma filament with a hot electron population is allowed to expand into the surrounding gas with density equal to the ion density inside the filament. All collisional processes are turned off, such that atoms are ionized only via the tunneling mechanism~\cite{Ammosov1986}. The Ar plasma filament is initially singly ionized with ion density $7 \times 10^{18}$ cm$^{-3}$ and radius 50 $\mu$m. The initial electron momentum distribution is isotropic and it has two components: a cold component with $p_e = 0.01$ $m_e c$ and a hot component with $p_e = m_e c$. We consider four cases where the hot component is 10\%, 30\%, 50\%, and 70\% of the initial population. Since the tunneling ionization rate is exponentially sensitive to the intensity of the applied electric field, we estimate, and confirm in simulations, that the sheath field can only ionize the surrounding gas to $Z=1$. We initialize a singly ionized filament in order to avoid a sharp discontinuity in $Z$ that is produced inside filaments with $Z > 1$ in our setup after the expansion begins. 

Snapshots of the electron density after the expansion and the radial electric field during the expansion for the 70\% case are shown in Fig.~\ref{Figure55}. The sheath electric field generated by the hot electrons exceeds 30 GV/m and can easily ionize the gas surrounding the plasma. This leads to a rapid increase of the plasma radius, which more than doubles over 1.5 ps. The maximum and angle-averaged plasma radius are comparable during the first 600 fs, indicating that the radial expansion is uniform. The instability that develops after 600 fs causes the expansion to continue in the form of radially protruding structures, leading to a considerable discrepancy between the maximum and average plasma radius. Figure~\ref{Figure55} also shows the maximum and angle-averaged radial velocity for this run, which are both in the range of 0.3~$c$ during the first 600 fs. The maximum velocity increases up to 0.6~$c$ with the onset of the instability. As we decrease the hot electron fraction, the expansion becomes slower and the plasma does not extend as far radially. This is consistent with the trends observed by decreasing the backing pressure from 10.5 to 7 bar, which leads to smaller clusters and effectively decreases the number of hot electrons. 

\begin{figure}[tb]
\centering
\subfigure{\includegraphics[scale=0.45]{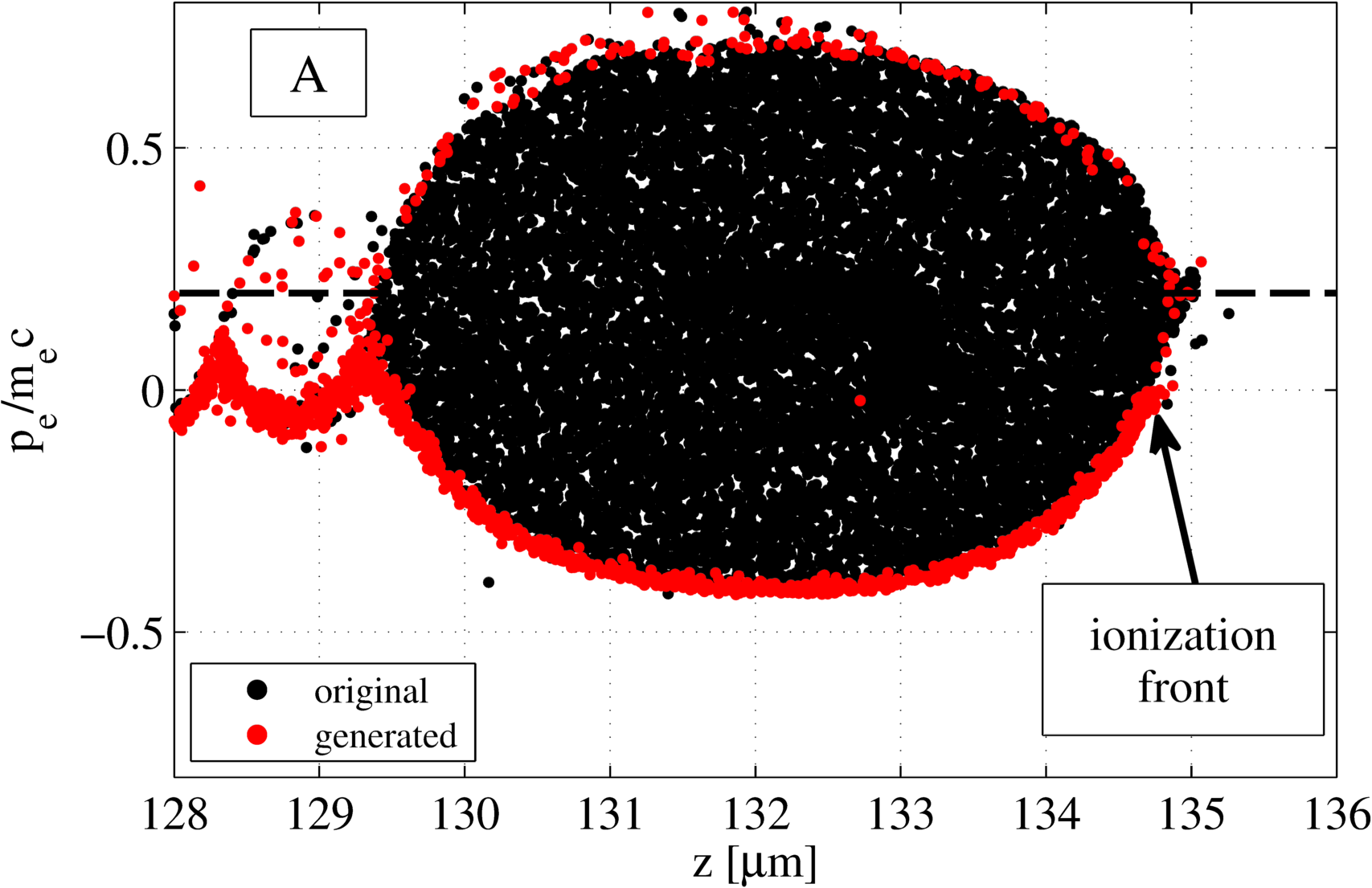}} \\
\subfigure{\includegraphics[scale=0.45]{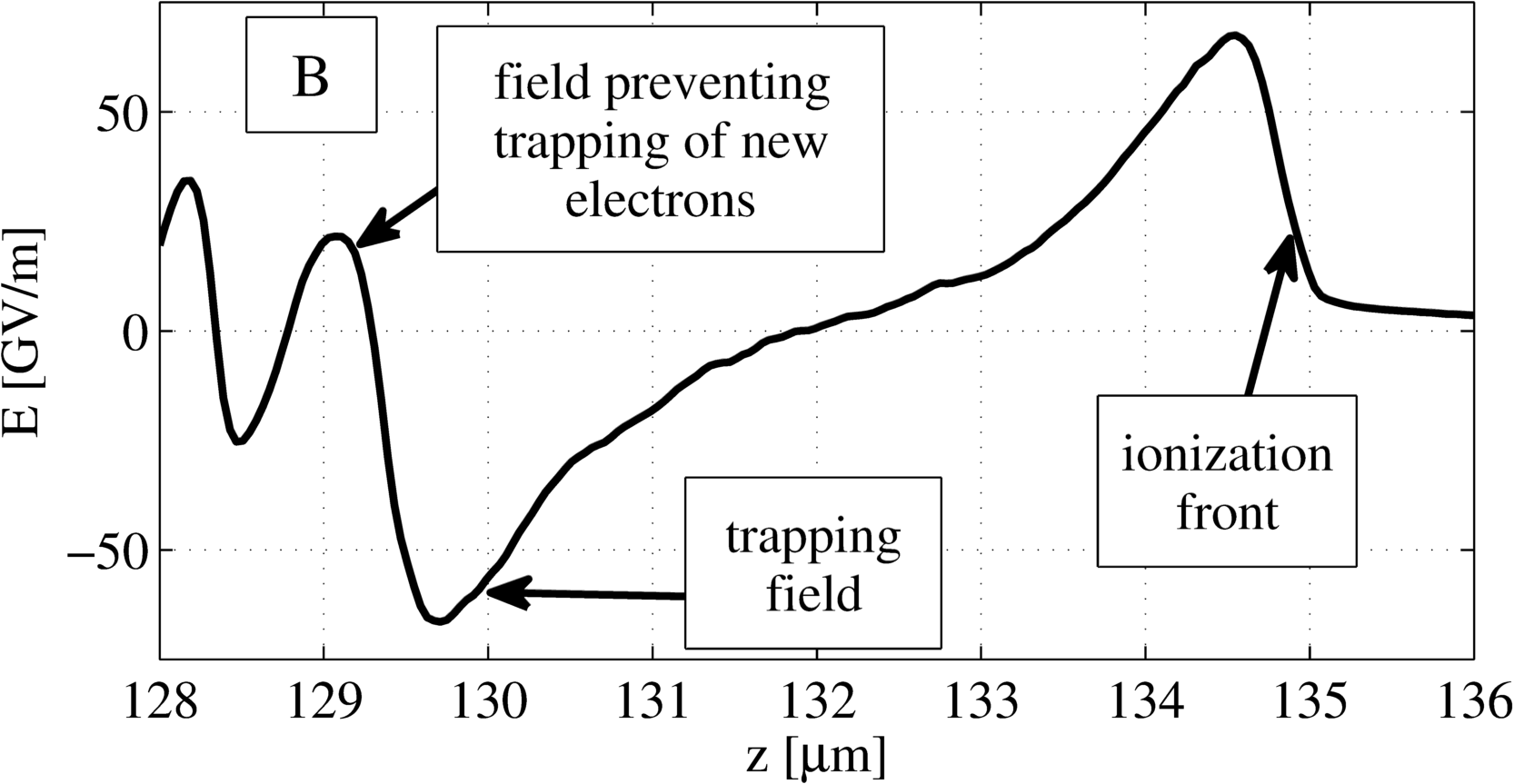}} \\
\subfigure{\includegraphics[scale=0.45]{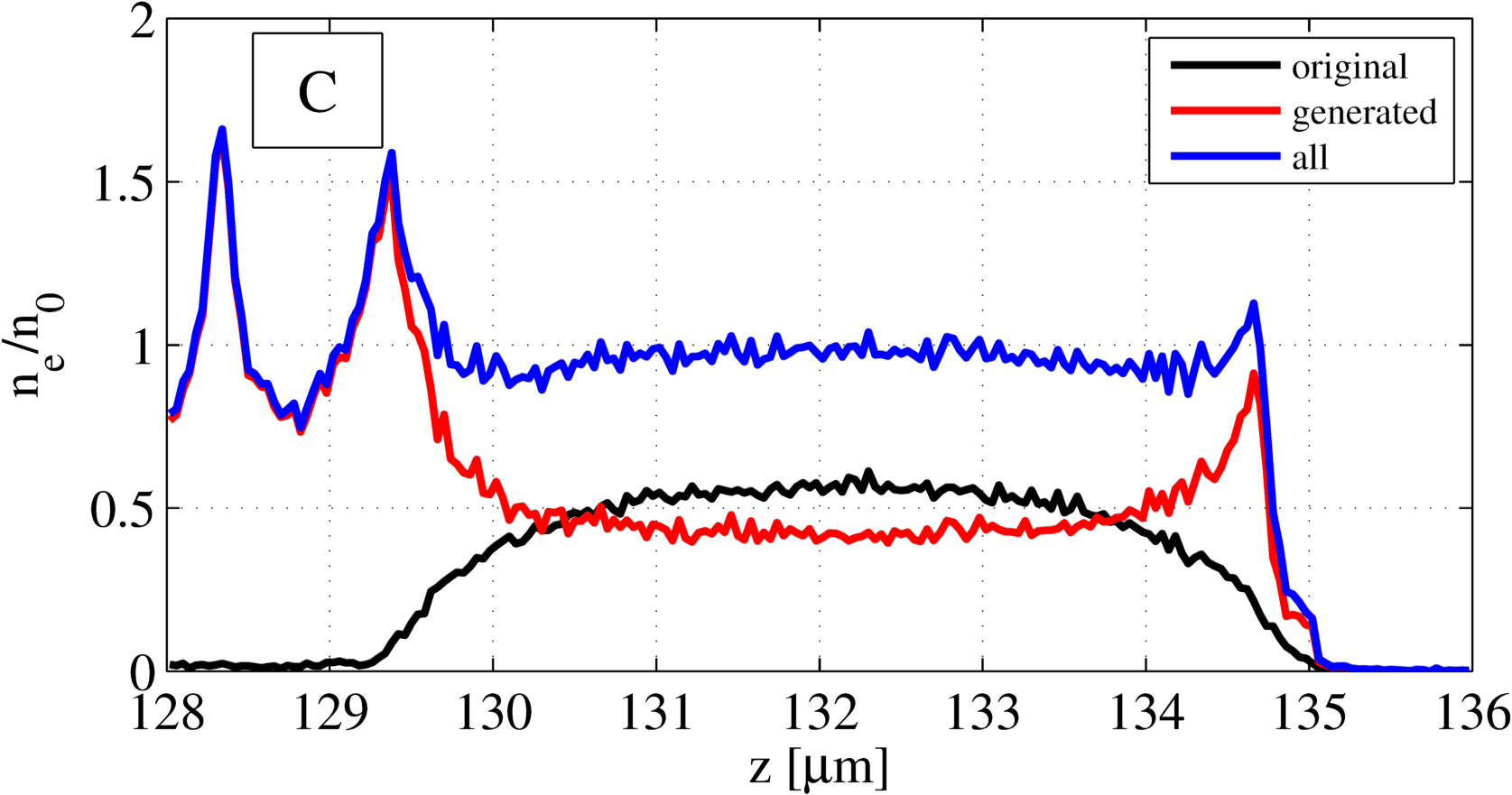}}
\caption{(Color online) Snapshots of electron phase-space, electric field, and electron densities in a 1D PIC simulation.}
\label{Figure_1D}
\end{figure}

The ionization in the simulations is caused exclusively by the sheath field, which raises the question of how this field can be sustained during the significant plasma expansion. One might expect for the interior of the plasma filament to serve as a potential well, with the electrons bouncing from edge to edge. If this were the case, the electron energy density $\varepsilon$ would drop as $1/R^2$ as the plasma expands. The sheath field would then drop as well, since it is proportional to $\sqrt{\varepsilon}$, and the expansion would stop before $R$ can increase significantly due to the exponential sensitivity of the tunneling ionization rate to the field strength. The image of the radial electric field in Fig.~\ref{Figure55} shows that a strong negative electric field develops just to the inside of the layer with a positive sheath field. Such field can prevent the hot electrons from spreading over the entire plasma, keeping them bunched near the boundary and allowing them to maintain a nearly constant strong sheath field.

In order to determine the mechanism generating the confining field, and whether it can lead to trapping of hot electrons, we have performed a 1D PIC simulation with a setup similar to that in 2D. The electron phase space is better visualized for a water-bag initial electron distribution with cut-offs at $p_e = \pm m_e c$ and hydrogen gas $(n_0 = 3.4 \times 10^{19}$ cm$^{-3})$. The initial ion density edge was located at 100~$\mu$m. Figure~\ref{Figure_1D} shows the right edge of an expanding plasma slab, including the electron momenta in the frame of the slab. Consider now the electron dynamics in a frame moving with the ionization front, where the sheath field remains static. The momentum corresponding to this frame is shown in Fig.~\ref{Figure_1D}a with a dashed line and, clearly, the new electrons born at the ionization front are more energetic than the electrons creating the ionizing field. The newly born electrons are accelerated away from the front by the sheath field. Their density drops below the ion density as a result of the flux conservation, producing a counteracting electric field (Fig. \ref{Figure_1D}b). The original electrons are less energetic in the moving frame, so they become trapped by this field while the generated electrons continue moving away. The counteracting field eventually slows down the generated electrons, and their density spikes. However, the resulting positive field prevents these electrons from reversing direction, so that they do not fully stop in the moving frame and never become trapped. These electrons carry away a small portion of the energy, which leads to a gradual decrease of the sheath field and the slowdown of the wave front.

In summary, we have presented measurements of a relativistic velocity ionization wave launched in an Ar jet containing clusters after it is irradiated by an intense laser beam. The measured velocity of the wave after the laser has passed is as high as 0.5 $c$ and the wave is sustained for up to 2 ps, causing the plasma radius to double. Such rapid ionization is caused by a sheath electric field at the edge of the plasma, which is generated by a hot electron population originated from laser-ionized clusters. The longevity of the wave is explained by a moving field structure that traps the hot electrons near the boundary, keeping them bunched and allowing them to maintain a strong sheath field despite the significant plasma expansion. A similar mechanism might be responsible for the rapid ionization observed in Kr and Xe clustering jets, where the inferred expansion velocity was 0.3 $c$~\cite{Gumbrell2008}. The instability observed in the 2D simulations results from azimuthal electron motion that causes fluctuations of the sheath field and exponential dependence of the ionization rate on the strength of the sheath field. A quantitative comparison of experimental data and simulations requires modeling of the laser interaction with an ensemble of clusters in order to determine more accurately the plasma conditions generated by the laser.

The simulations were performed using Plasma Simulation Code~\cite{PSC} using HPC resources provided by the Texas Advanced Computing Center at The University of Texas. This work was supported by NNSA Contract No. DE-FC52-08NA28512, U.S. DOE Contracts No. DE-FG02-04ER54742, and a NPSC fellowship awarded to M. M.

\end{document}